\documentclass[conference]{IEEEtran}
\IEEEoverridecommandlockouts

\usepackage{cite}
\usepackage{amsmath,amssymb,amsfonts}
\usepackage{algorithmic}
\usepackage{graphicx}
\usepackage{textcomp}
\usepackage{xcolor}
\def\BibTeX{{\rm B\kern-.05em{\sc i\kern-.025em b}\kern-.08em
    T\kern-.1667em\lower.7ex\hbox{E}\kern-.125emX}}
\usepackage{booktabs} 
\usepackage{multirow}
\usepackage{hyperref}

\begin{document}

\title{Latent Space Disentanglement via Activation Steering for Interpretable Attribute Control in Symbolic Music Generation\\
}

\author{\IEEEauthorblockN{Ioannis Prokopiou}
\IEEEauthorblockA{\textit{Athens University of Economics and Business}\\
\textit{Orfium} \\
Athens, Greece \\
gian.prokopiou@aueb.gr}

\and
\IEEEauthorblockN{Pantelis Vikatos}
\IEEEauthorblockA{\textit{Innovation Lab} \\
\textit{Orfium}\\
Athens, Greece \\
pantelis@orfium.com}
\and
\IEEEauthorblockN{Maximos Kaliakatsos-Papakostas}
\IEEEauthorblockA{\textit{Department of Music Technology and Acoustics } \\
\textit{Hellenic Mediterranean University}\\
Rethymno, Greece \\
maximoskp@hmu.gr}
\and
\IEEEauthorblockN{Theodoros Giannakopoulos}
\IEEEauthorblockA{\textit{Institute of Informatics \& Telecommunications} \\
\textit{National Center for Scientific Research “Demokritos”}\\
Athens, Greece \\
 tyianak@iit.demokritos.gr}
\and
\IEEEauthorblockN{Themos Stafylakis}
\IEEEauthorblockA{\textit{Department of Informatics} \\
\textit{Athens University of Economics and Business}\\
\textit{Archimedes/Athena R.C}\\
Athens, Greece \\
tstafylakis@aueb.gr}
}

\maketitle

\begin{abstract}
Transformer-based architectures have significantly advanced the generation of complex symbolic sequences, yet a significant gap remains in achieving fine-grained, interpretable control over discrete signal attributes. This paper investigates the mechanistic interpretability of the Multitrack Music Transformer (MMT) and proposes a framework for deterministic attribute modulation without retraining to bridge this gap via inference-time activation steering. Utilizing the Difference-in-Means (DiffMean) methodology, we isolate latent directions for signal attributes, specifically Pitch and Duration, within the residual stream. We validate the Linear Representation Hypothesis in this domain, achieving high correlation between steering magnitude and attribute shift. To address the inherent feature entanglement in multi-attribute steering, we introduce a Dual Steering framework utilizing Gram-Schmidt Orthogonalization.  Experimental results demonstrate that this geometric decoupling reduces conceptual interference and signal degradation compared to naive vector addition, enabling independent deterministic control even against strong autoregressive conditioning.
\end{abstract}


\section{Introduction}
The integration of deep learning into the creative arts has produced state-of-the-art autoregressive models capable of respecting long-term structural dependencies \cite{anantrasirichai2022artificial}. However, these models often function as ``black boxes'' \cite{tredinnick2023black} where the relationship between high-dimensional internal activations and discrete output attributes is obscured. For applications requiring precise intervention such as algorithmic composition or signal synthesis, global conditioning often lacks the granularity needed for specific manipulation of internal states. There is a pressing need for precise, inference-time intervention methods that do not require computationally expensive retraining without disrupting the entire generative process \cite{li2023inference}. 

The field of mechanistic interpretability \cite{bereska2024mechanistic} seeks to reverse-engineer neural networks to understand the algorithms they implement. A central tenet is the Linear Representation Hypothesis, which posits that high-level concepts are represented as vectors within the activation space \cite{elhage2021toy, Zou2023RepresentationEA}. While validated in Large Language Models (LLMs), its application to symbolic signal domains remains nascent. Identifying these vectors enables Activation Steering, the process of intervening on the model's internal activations during the forward pass to bias the generation toward a desired attribute \cite{rimsky2024steering}. 

While recent advancements have led to unified frameworks capable of high-quality generation via multi-modal conditioning, such as Seed-Music \cite{Bai2024SeedMusicAU}, XMusic \cite{11091494}, and the flow-matching based JASCO \cite{Tal2024JointAA}, these systems often require extensive retraining or complex architectural changes. This work addresses the problem of inference-time steerability. Models such as the Multitrack Music Transformer (MMT) \cite{dong2023multitrack} utilize specialized data representations to encode discrete musical events into token sequences, learning the high-dimensional probability distributions of musical corpora like the Symbolic Orchestral Database (SOD) \cite{crestel2017database}. Rather than focusing on computationally expensive fine-tuning or prompt engineering, we apply mechanistic interpretability to reverse-engineer the latent representations of a pre-trained MMT, focusing on Average Note Pitch and Average Note Duration. 

Our contribution provides a layer-wise sensitivity analysis to map attribute encoding topology and investigate linear steerability within the MMT and evaluates steering dynamics across the network depth. We implement orthogonalization methods to disentangle correlated features, ensuring independent modulation and precise dual-concept control. Audio examples and code on \url{https://giannisprokopiouorfium.github.io/music-transformer-sae/}.

\section{Related Work}
Symbolic music generation has transitioned from rule-based systems to sophisticated deep learning architectures, with the Transformer  \cite{vaswani2017attention} architecture revolutionising the modeling of long-term structural dependencies. The MMT introduced a compact 6-tuple event representation, incorporating type, beat, position, pitch, duration, and instrument, to significantly reduce sequence lengths compared to prior formats like REMI+ \cite{von2022figaro}. This efficiency makes MMT an ideal backbone for real-time steering research. 

Activation engineering \cite{turner2025steering} facilitates this shift by optimizing internal model states rather than input prompts grounded in the Linear Representation Hypothesis \cite{10.5555/3692070.3693675}. Research on LLMs has demonstrated that injecting these vectors during the forward pass can deterministically bias outputs without weight optimization \cite{kang2025model}. The Difference-in-Means (DiffMean) \cite{Marks2023TheGO, rimsky2024steering} methodology serves as a robust baseline for extracting these vectors by capturing the centroid difference between contrasting concept clusters. Recent benchmarks like AxBench \cite{wu2025axbench} confirm that DiffMean often outperforms complex alternatives like Sparse Autoencoders (SAEs) for steering tasks.

In the musical domain, mechanistic interpretability \cite{bereska2024mechanistic} is an emerging frontier. Early attempts at controllable synthesis utilized latent regularization in architectures like Music FaderNets \cite{tan2020music}, or compositional augmentation of Transformer architectures \cite{young2022compositional}. Panda et al. \cite{panda2025fine} demonstrated that steering residual streams and attention heads can enable fine-grained style transfer and genre fusion. Modern techniques such as SMITIN \cite{koo2025smitin} employ self-monitored interventions with classifier probes to dynamically adjust steering strength, while MusicRFM \cite{Zhao2025SteeringAM} analyzes internal gradients to identify concept directions for autoregressive models. Most relevant to our study is the work of Facchiano et al. \cite{Facchiano2025ActivationPF}, which applies activation patching to interpretable latent directions, confirming that localized interventions can effectively modulate high-level musical attributes. In their work they applied activation patching to MusicGen \cite{10.5555/3666122.3668188} to control tempo and timbre, identifying mid-range layers as the most effective locus for intervention. Our research distinguishes itself by applying these concepts to the symbolic domain, where the discrete nature of MIDI-based attributes allows for mathematically precise evaluation and manipulation.

A persistent challenge in multi-attribute steering is the conceptual interference between correlated features \cite{yao2026large}, such as the natural entanglement of pitch and duration. While techniques involving geometric operations, such as angular steering \cite{vu2025angular}, have been proposed to define stable steering planes in latent space, and the Composer Vector method \cite{composervector2025} that utilizes latent space arithmetic to fuse styles, we build upon these mathematical foundations by applying Gram-Schmidt Orthogonalization \cite{bjorck1994numerics}. By explicitly decoupling correlated features, this approach prevents interference during generation and addresses the entanglements typically learned by the model's residual stream, ensuring independent control over discrete musical attributes.

\section{Methodology}

\subsection{Signal Representation and Model Infrastructure}
We utilize the pre-trained MMT, a decoder-only Transformer architecture trained on the SOD dataset. The model processes musical events as discrete 6-dimensional tuples, which are embedded into a 512-dimensional continuous vector space. Each event tuple consists of:
\begin{itemize}
    \item \textbf{Type (4 tokens):} Categorizes the event (note, time-shift, end-of-song, instrument-change).
    \item \textbf{Beat (257 tokens):} Integer beat position (0 to 256).
    \item \textbf{Position (384 tokens):} Sub-beat resolution at intervals.
    \item \textbf{Pitch (128 tokens):} MIDI values ranging from 0–127.
    \item \textbf{Duration (768 tokens):} Note length in ticks, from staccato (1 tick) to sustained (768).
    \item \textbf{Instrument (128 tokens):} MIDI program numbers for orchestral instrumentation (0 to 127).
\end{itemize}
Our steering interventions target Pitch (dimension 4) and Duration (dimension 5) for interpretability by modifying the residual stream activations that influence the probability distribution over these discrete vocabularies at generation time.

\subsection{Latent Vector Extraction and Concept Definition}
To compute steering vectors via DiffMean, we defined “High” and “Low” concept thresholds by analyzing the statistical distribution of the SOD corpus. Thresholds (Table \ref{tab:data_curation}) were empirically set at the 20th and 80th quantiles, yielding robust, representative, non-overlapping clusters and stable latent directions across alternative extreme quantile choices.

\begin{table}[t]
\centering
\caption{Data Curation Quintile Thresholds for Vector Extraction.}
\label{tab:data_curation}
\begin{tabular}{lccc}
\hline
\textbf{Concept} & \textbf{Metric} & \textbf{Low (20\%)} & \textbf{High (80\%)} \\
\hline
Note Pitch & Avg MIDI Number & 60 & 67.6 \\
Note Duration & Avg MIDI Ticks & 6.5 & 14.5 \\
\hline
\end{tabular}
\end{table}

\begin{figure}[htbp]
    \centering
    \includegraphics[width=0.7\linewidth]{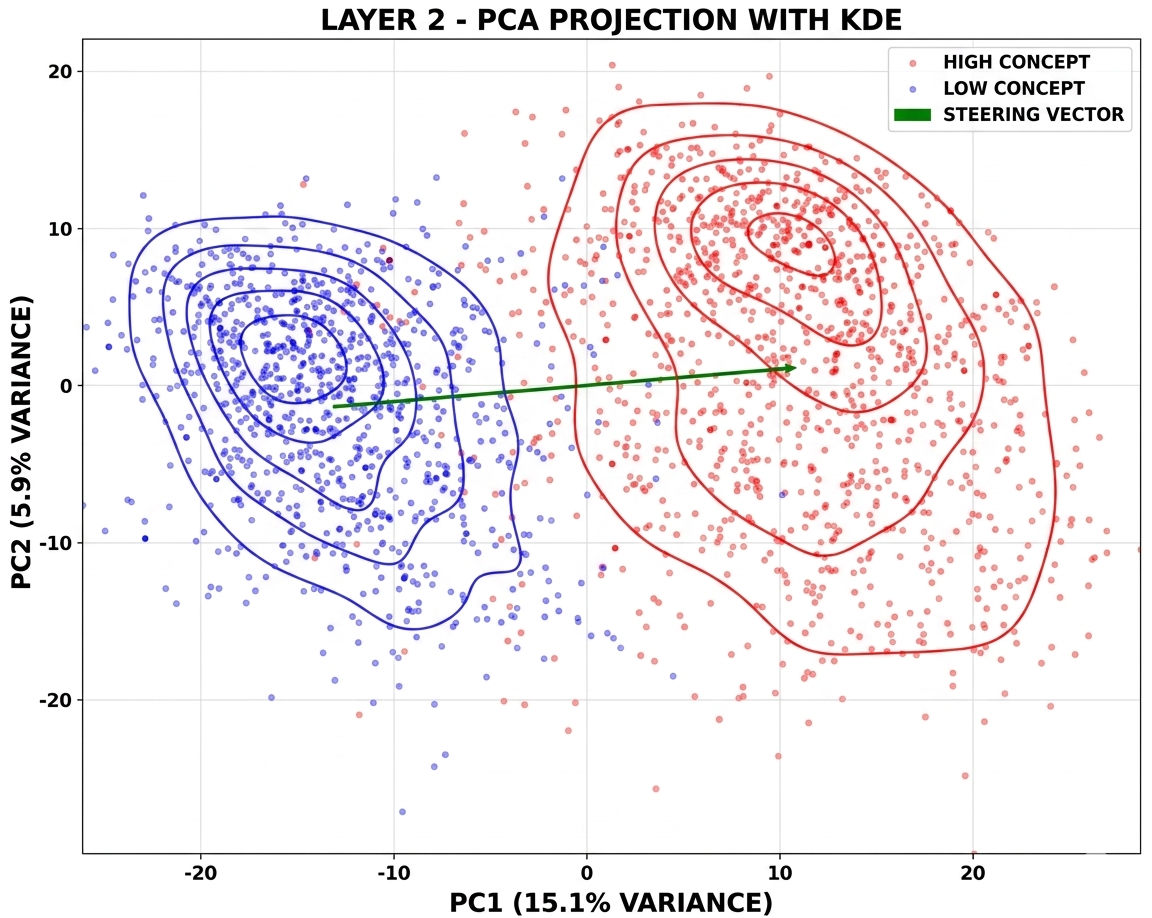} 
    \caption{\textbf{Latent Representation of Note Duration at Layer 2.} A 2D Kernel Density Estimation of the PCA-projected activations in the MMT residual stream. Clear clustering of long/short duration tokens in early layers validates the linear separability of rhythmic features prior to melodic processing.}
    \label{fig:layer_2_duration_kde}
\end{figure}

We intercept the residual stream at the output of every transformer decoder block using forward hooks as it contains the model's semantic musical representations, rather than the output layer that only projects to vocabulary space for next-token prediction. For a given layer $l$, we extract the summary activation $h$ corresponding to the last valid token that has full sequence context. The steering vector $v^{(l)}$ is calculated as shown in \eqref{eq:steering_vector} as the difference between the means of the high-attribute and low-attribute set. Here, $N_{pos}$ and $N_{neg}$ denote the number of samples in the high and low concept clusters balanced at $1,280$ samples each based on the previously defined quintile thresholds, while $h(x)^{(l)}$ represents the activation vector for a specific input segment $x$ at layer $l$.

\begin{equation}
\label{eq:steering_vector}
v^{(l)} = \frac{1}{N_\text{pos}} \sum_{i=1}^{N_\text{pos}} h(x_\text{pos, i})^{(l)} - \frac{1}{N_\text{neg}} \sum_{j=1}^{N_\text{neg}} h(x_{\text{neg}, j})^{(l)}
\end{equation}
Our sensitivity analysis reveals that Note Pitch is most linearly separable at Layer 11, while Note Duration shows peak separation at Layer 2 shown in Fig. \ref{fig:layer_2_duration_kde}, suggesting that rhythmic features are encoded earlier in the network than melodic ones.

\subsection{Inference-Time Generation Steering}
To steer the model's behavior during inference, we modify the hidden states $h^{(l)}$ by injecting the steering vector $v^{(l)}$ scaled by a coefficient $\alpha$ as shown in \eqref{eq:steering}:
\begin{equation}
\label{eq:steering}
h_\text{steer}^{(l)} \leftarrow h^{(l)} + \alpha v^{(l)}
\end{equation}
We evaluated several injection strategies: \textit{All-to-All} (layer-specific vectors injected at every layer), \textit{One-to-All} (broadcasting a single optimal direction), and \textit{Some-to-Some} (targeting sensitive sub-layers). Empirical results indicate that the All-to-All strategy yields the most robust trade-off between steering potency and generation quality for the symbolic MMT.

\subsection{Dual Steering and Orthogonal Disentanglement}
A critical challenge in multi-attribute steering is conceptual interference; for example, high-pitch passages in the training corpus often statistically correlate with shorter durations. We measured an average absolute cosine similarity of $0.49$ between Pitch and Duration vectors across the network, peaking at $0.81$ in Layer 3. To enable independent manipulation and mitigate interference between features, we evaluated four strategies to construct the combined steering vector $v_\text{combined}$:

\begin{enumerate}
    \item \textbf{Simple Addition:} A baseline linear combination where vectors are added without modification in \eqref{eq:simple}:
    \begin{equation}
    \label{eq:simple}
        v_\text{combined} = \alpha_{p} v_{p} + \alpha_{d} v_{d}
    \end{equation}
    
    \item \textbf{Gram-Schmidt (Pitch Priority):} To preserve pitch purity, the duration vector is orthogonalized against the pitch vector ($v_{d}^{\perp}$) shown in \eqref{eq:grahm}:
    \begin{equation}
    \label{eq:grahm}
        v_{d}^{\perp} = v_{d} - \frac{v_{d} \cdot v_{p}}{\|v_{p}\|^2} v_{p} \quad \Rightarrow \quad v_\text{combined} = \alpha_{p} v_{p} + \alpha_{d} v_{d}^{\perp}
    \end{equation}
    
    \item \textbf{Gram-Schmidt (Duration Priority):} Conversely, we prioritize rhythmic stability by orthogonalizing the pitch vector against duration ($v_{p}^{\perp} = v_{p} - \text{proj}_{v_{d}}(v_{p})$).
    
    \item \textbf{Symmetric Orthogonalization:} We utilize Singular Value Decomposition (SVD) ~\cite{shah2006symmetry} to define a subspace where both vectors are mutually orthogonal, maximizing simultaneous independence.
\end{enumerate}

\section{Experimental Evaluation}

To rigorously assess the steerability of symbolic musical attributes, we conducted experiments across two paradigms: 

\begin{itemize}
    \item \textbf{Unconditional Generation:} Sequences are generated from an empty context to isolate the steering vector's influence on the model’s intrinsic priors. This measures the coherence and structural integrity when governed solely by pretrained weights and the injected vector.
    \item \textbf{Conditional Generation:} To evaluate the steering vector’s capacity to override strong local contexts, we rank SOD training tracks by average pitch or duration in the first 16 beats and select \textit{extreme} cases. These 16 beats are used as a conditioning prefix to establish a statistically strong autoregressive prior, after which we attempt to steer the continuation in the opposite direction (e.g., imposing low pitch following a high-pitch context).
\end{itemize}



Performance is quantified via \textbf{Steering Success}, percentage of generations successfully moving in the intended direction, and \textbf{Quality Degradation ($\delta$)}, the cumulative absolute deviation from SOD baselines: Pitch Class Entropy (2.974), Scale Consistency (92.26\%), and Groove Consistency (93.05\%). These metrics are inherently weighted toward pitch-related factors and thus serves as a proxy for human perception, $\delta$ correlates directly with melodic integrity; a $\delta \approx 10$ indicates severe dissonance akin to random note generation, whereas $\delta \le 3$ remains perceptually coherent to listeners. Generations used temperature $1.0$ and Top-K filtering (\textit{filter\_thres} $= 0.9$).


\subsection{Single-Attribute Unconditioned Dynamics}

\subsubsection{Pitch Steering Dynamics}
Unconditioned pitch steering trials revealed a robust and predictable linear response with $R^2 = 0.8154$. Statistical analysis of 50 generations across the alpha grid between $-2$ and $2$ produced a Pearson correlation coefficient of $r = 0.9030$ ($p = 0.0009$), indicating that roughly $82\%$ of the variance in output pitch is directly attributable to steering strength. The linear slope was estimated at $+13.35$ semitones per unit of $\alpha$. As shown in Table \ref{tab:attribute_control}, the effect demonstrated asymmetry regarding the baseline mean of $65.73$: positive steering ($\alpha = +2.0$) shifted the mean pitch by $+15.50$ semitones ($+23.6\%$), whereas negative steering ($\alpha = -2.0$) induced a shift of $-29.00$ semitones ($-44.1\%$). 

\subsubsection{Duration Steering}
Duration steering in the same configuration exhibited a monotonic relationship with a Pearson correlation of $r = 0.9263$ ($p = 0.0003$) and linear response with $R^2 = 0.8580$. The linear model suggests a rate of change of approximately $+10.77$ ticks per $\alpha$. Response was asymmetric; negative $\alpha$ values hit a physical lower at $\sim 3.08$ ticks (a $59.0\%$ reduction), positive $\alpha$ values produced  expansions of up to $+406.9\%$ ($\alpha = +2.0$) as shown in Table \ref{tab:attribute_control}. 

\begin{table}[t]
\centering
\caption{Single-Attribute Unconditioned Steering using $|\alpha|=2.0$.}
\setlength{\tabcolsep}{3pt}
\begin{tabular}{l l c c c}
\toprule
\textbf{Attribute} & \textbf{Variant} & \textbf{Absolute} & \textbf{Relative} & \textbf{Quality Degradation} \\
\midrule
\multirow{3}{*}{Pitch} & MMT (Base) & $65.73 \pm 10.10$ & -- & -- \\
 & $\rightarrow$ low & $36.73 \pm 3.81$ & 44.1\% & 0.25 \\
 & $\rightarrow$ high & $81.23 \pm 6.32$ & 23.6\% & 2.01 \\
\midrule
\multirow{3}{*}{Duration} & MMT (Base) & $7.51 \pm 4.59$ & -- & -- \\
 & $\rightarrow$ short & $3.08 \pm 0.51$ & 59.0\% & 0.90 \\
 & $\rightarrow$ long & $38.06 \pm 20.88$ & 406.8\% & 1.97 \\
\bottomrule
\end{tabular}
\label{tab:attribute_control}
\end{table}

\subsection{Single Context Override Contrasting Scenarios}

To rigorously evaluate the robustness of our steering vectors, we designed experimental contrasting scenarios to test whether internal activations can override strong autoregressive context. We utilize a 16-beat conditioning prefix from training samples exhibiting extreme attribute values and generate 512-token continuations while injecting steering vectors across a comprehensive grid of strengths ($\alpha \in \{\pm 0.5, \pm 0.75, \pm 1.0, \pm 1.25, \pm 1.5\}$). Alphas were filtered per scenario to test only directionally-relevant combinations (e.g., negative $\alpha$ for High$\to$Low transitions).

\subsubsection{Pitch Context Override}
We evaluated \textbf{Low$\to$High} (conditioning pitch 32.9--47.8 MIDI) and \textbf{High$\to$Low} (conditioning pitch 68.4--82.7 MIDI) scenarios. Vectors achieved a $93.4\%$ overall success rate with an average magnitude shift of $31.7 \pm 4.3$ semitones ($\sim$2.6 octaves). We observed a Directional Asymmetry where upward steering was more effective ($96.1\%$ success) than downward ($85.6\%$), suggesting a model bias toward lower pitch ranges. Optimal success was achieved at $\alpha_{pitch} = +0.75$ for upward success with $0.02$ degradation and $-1.25$ for downward with $0.05$ degradation.

\subsubsection{Duration Context Override}
Duration tests evaluated \textbf{Short$\to$Long} (conditioning duration 1.87--6.40 ticks) and \textbf{Long$\to$Short} (conditioning duration 14.89--17.0 ticks) transitions, an overall success rate of $91.7\%$ with an average magnitude shift of $12.1 \pm 1.2$ ticks. For positive $\alpha$, Short$\to$Long reached $97.8\%$ success at $\alpha_{dur} = +0.75$. For negative $\alpha$, Long$\to$Short achieved $99.4\%$ success at $\alpha_{dur} = -1.0$, bounded by the 3-tick floor. Degradation for expansion ($\delta$: $1.8$--$3.4$) scaled sub-linearly, while compression degradation scaled linearly ($2.1$--$4.3$), suggesting the model’s internal representations are optimized for moderate-to-long durations.

\subsection{Layer Sensitivity and Injection Strategy}
To identify the optimal locus of intervention, we conducted systematic grid searches across the transformer's depth, comparing \textbf{All-to-All} (systemic bias), \textbf{One-to-All} (single broadcast), and \textbf{Some-to-Some} (targeted groups). For Pitch, Layer 10 was most potent for reduction, while Layer 11 was optimal for increase. For Duration, later layers (Layers 8--11) exhibited the highest sensitivity. Empirical knowledge across both attributes shows that while specific layers provide targeted control for either decrease or increase, the \textbf{All-to-All} strategy remains the most robust intervention paradigm. By introducing a systemic bias in the residual stream, it achieved the superior trade-off, providing balanced control in either increase or decrease nudging the global context for both attributes with minimal degradation compared to the localized strategies.


\subsection{Dual Steering and Disentanglement Results}

\subsubsection{Unconditioned Strategy Validation}

\begin{figure*}[t] 
    \centering
    \includegraphics[width=0.85\textwidth]{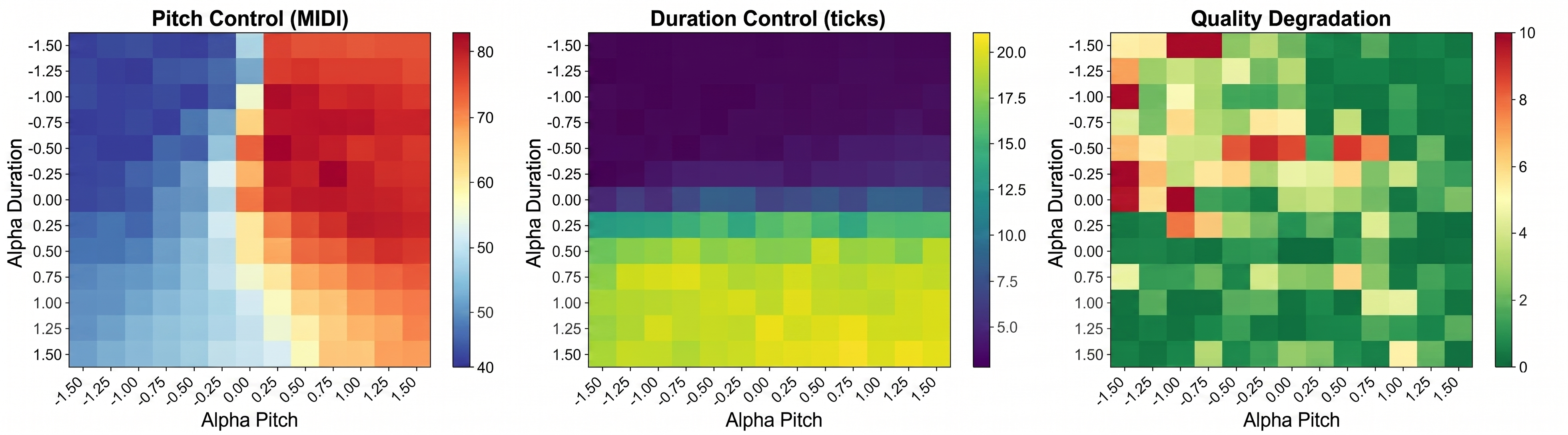}
    \caption{\textbf{Dual Steering Grid Search Heatmaps.} Heatmap of quality degradation across alpha combinations.}
    \label{fig:triple_heatmap}
\end{figure*}

Using a validation grid of $1,600$ generations, we compared four composition strategies to address conceptual interference. Success is measured via the Dual Steering Success and Quality Degradation ($\delta$). Results in Table \ref{tab:dual_results} show that Gram-Schmidt (Pitch Priority) outperformed other methods, achieving 88.5\% success with the lowest degradation ($\delta=2.14$). In contrast, Symmetric Orthogonalization via SVD, implemented by extracting the scaled orthonormal basis from the $U$ and $S$ matrices of the stacked vector matrix $V=USV^T$, underperformed at 79.3\%. This suggests that symmetric rotation dilutes both concept directions, whereas Gram-Schmidt preserves the primary pitch vector's integrity. This supports an inherent hierarchy where pitch serves as a fundamental anchor. As Fig. \ref{fig:triple_heatmap} shows, each $alpha$ steers the expected corresponding features while showing an optimal configuration at moderate positive alphas (0.75--1.25 range) where dual steering effectiveness is maximized with minimal conceptual interference.

\begin{table}[t]
\centering
\caption{Uncoditioned Dual Steering Strategy Comparison}
\label{tab:dual_results}
\begin{tabular}{lcc}
\hline
\textbf{Strategy} & \textbf{Dual Steering Success} & \textbf{Quality Degradation} \\
\hline
\textbf{Gram-Schmidt (Pitch)} & \textbf{88.5\%} & \textbf{2.14} \\
Simple Addition & 85.2\% & 2.31 \\
Gram-Schmidt (Dur) & 82.7\% & 2.45 \\
Symmetric Orthogonal & 79.3\% & 2.68 \\
\hline
\end{tabular}
\end{table}

\subsubsection{Conditioned Dual Steering}
For conditioned dual steering, we forced the model to counteract contradictory 16-beat conditioning context from training samples that were fell into both categories calibrated thresholds using a 10$\times$10 alpha grid ($\alpha \in \{\pm 0.5, \pm 0.75, \pm 1.0, \pm 1.25, \pm 1.5\}$) that was filtered per scenario to test only directionally relevant combinations under Gram-Schmidt Orthogonalization (Pitch Priority) with All-to-All layer injection. The results are shown in Table \ref{tab:fight_scenarios} for the different contrasting scenarios.

\begin{table}[t]
\centering
\caption{Dual Steering Performance by Conditioned Scenario}
\label{tab:fight_scenarios}
\begin{tabular}{lcccc}
\toprule
\textbf{Scenario} & \textbf{Success} & \textbf{Avg $\delta$} & \textbf{Avg $|\Delta P|$} & \textbf{Avg $|\Delta D|$} \\
\midrule
Low/Short $\to$ High/Long & 96.1\% & 3.03 & 28.6 ST & 12.8 T \\
Low/Long $\to$ High/Short & 90.6\% & 1.33 & 27.8 ST & 11.1 T \\
High/Long $\to$ Low/Short & 85.6\% & 2.59 & 35.4 ST & 13.4 T \\
High/Short $\to$ Low/Long & 82.2\% & 5.80 & 35.3 ST & 12.0 T \\
\bottomrule
\end{tabular}
\end{table}

Emprirically, the best configuration for dual steering contains $\alpha_{pitch} \in [\pm 0.5, \pm 1.0]$ and $\alpha_{dur} \in [\pm 0.5, \pm 0.75]$. Beyond $|\alpha| > 1.25$, degradation increases exponentially. 
Key insights include:
\begin{enumerate}
    \item \textbf{Directionality Asymmetry:} Upward steering is significantly easier ($96.1\% success$) than downward ($82.2\%$), reflecting the training distribution's melodic bias.
    \item \textbf{Quality Cost:} The High/Short $\to$ Low/Long scenario exhibited $2.8\times$ higher degradation than its inverse (5.80 vs 1.33), indicating high costs for fighting extreme high-pitch contexts.
    \item \textbf{Context Dominance:} Steering successfully overrode local context in $88.6\%$ of trials on average, with effects (2--3 octaves) far exceeding typical transposition ranges.
\end{enumerate}

\section{Conclusions}
This study explores the Linear Representation Hypothesis in symbolic music, confirming that the MMT encodes attributes such as pitch and duration as linear latent directions, which can be used for activation steering as a training-free method for precise musical control. The systemic All-to-All injection strategy provides the most robust steering, successfully overriding extreme autoregressive contexts. By applying Gram-Schmidt Orthogonalization, we decoupled entangled features to enable novel combinations. Future work will utilize SAEs, adaptive or feedback-based control and reverse-prompting to isolate monosemantic features for abstract concepts.

\section*{Acknowledgment}
This research was funded by the European Union’s Horizon Europe research and innovation programme under the AIXPERT project (Grant Agreement No. 101214389), which aims to develop an agentic, multi-layered, GenAI-powered framework for creating explainable and transparent AI systems.

\bibliographystyle{IEEEtran} \bibliography{references} 

\end{document}